\documentclass[%
reprint,
superscriptaddress,
showpacs,
floatfix, ]%
{revtex4-1}

\usepackage{color}

\usepackage{graphicx}
\usepackage{dcolumn}
\usepackage{bm}
\usepackage[dvipdfm,bookmarks=true,colorlinks,%
            citecolor=blue,linkcolor=blue,hypertex, %
            breaklinks=true]{hyperref}

\begin{document}

\title{Odd systems in deformed relativistic Hartree Bogoliubov theory in continuum}

\author{Lulu Li}
 \affiliation{State Key Laboratory of Nuclear Physics and Technology,
              School of Physics, Peking University, Beijing 100871, China}
\author{Jie Meng}
 \affiliation{State Key Laboratory of Nuclear Physics and Technology,
              School of Physics, Peking University, Beijing 100871, China}
 \affiliation{School of Physics and Nuclear Energy Engineering, Beihang University,
              Beijing 100191, China}
 \affiliation{State Key Laboratory of Theoretical Physics,
              Institute of Theoretical Physics, Chinese Academy of Sciences,
              Beijing 100190, China}
 \affiliation{Department of Physics, University of Stellenbosch, Stellenbosch, South Africa}
\author{P. Ring}
 \affiliation{Physikdepartment, Technische Universit\"at M\"unchen,
              85748 Garching, Germany}
 \affiliation{State Key Laboratory of Nuclear Physics and Technology,
              School of Physics, Peking University, Beijing 100871, China}
\author{En-Guang Zhao}
 \affiliation{State Key Laboratory of Theoretical Physics,
              Institute of Theoretical Physics, Chinese Academy of Sciences,
              Beijing 100190, China}
 \affiliation{State Key Laboratory of Nuclear Physics and Technology,
              School of Physics, Peking University, Beijing 100871, China}
 \affiliation{Center of Theoretical Nuclear Physics, National Laboratory
              of Heavy Ion Accelerator, Lanzhou 730000, China}
\author{Shan-Gui Zhou}
\email{sgzhou@itp.ac.cn}
 \affiliation{State Key Laboratory of Theoretical Physics,
              Institute of Theoretical Physics, Chinese Academy of Sciences,
              Beijing 100190, China}
 \affiliation{Center of Theoretical Nuclear Physics, National Laboratory
              of Heavy Ion Accelerator, Lanzhou 730000, China}

\date{\today}

\begin{abstract}
In order to describe the exotic nuclear structure in unstable odd-$A$ or
odd-odd nuclei, the deformed relativistic Hartree Bogoliubov theory
in continuum has been extended to incorporate the blocking effect
due to the odd nucleon.
For a microscopic and self-consistent description of pairing correlations,
continuum, deformation, blocking effects, and the extended spatial density
distribution in exotic nuclei, the deformed relativistic Hartree Bogoliubov
equations are solved in a Woods-Saxon basis in which the radial wave
functions have a proper asymptotic behavior at large $r$.
The formalism and numerical details are provided.
The code is checked by comparing the results with those of spherical
relativistic continuum Hartree Bogoliubov theory in the nucleus $^{19}$O.
The prolate deformed nucleus $^{15}$C is studied
by examining the neutron levels and density distributions.
\end{abstract}



\maketitle

As one of the most fascinating phenomena found in exotic
nuclei, the nuclear halo contains many new
and interesting features, such as a largely extended spatial density
distribution, dilute and pure neutron matter, the Borromean property,
and the coupling between bound states and the continuum~\cite{Tanihata1985_PRL55-2676}.
In the past, several relativistic Hartree Bogoliubov theories with or without
the Fock term have been developed for a self-consistent description of spherical
halo nuclei~\cite{Meng1996_PRL77-3963, Meng1998_PRL80-460, Meng1998_NPA635-3,
Poschl1997_PRL79-3841, Long2010_PRC81-024308}.
In order to investigate halo phenomena in deformed nuclei in a microscopic
and self-consistent way, a deformed relativistic Hartree Bogoliubov
(RHB) theory in continuum has been developed for even-even nuclei~\cite{Zhou2008_ISPUN2007,
Zhou2010_PRC82-011301R, Li2012_PRC85-024312}.
Within this model a decoupling in shape between the core and the halo
has been predicted in some deformed nuclei close to the neutron drip line,
e.g, in $^{42,44}$Mg~\cite{Zhou2010_PRC82-011301R, Li2012_PRC85-024312}.

In order to describe the exotic nuclear structure in unstable odd-$A$ or odd-odd
nuclei, the blocking effect of one or several nucleons has to be taken into account.
In the present Letter, the deformed relativistic Hartree Bogoliubov theory
in continuum~\cite{Zhou2008_ISPUN2007, Zhou2010_PRC82-011301R, Li2012_PRC85-024312}
is extended to incorporate the blocking effect due to an odd nucleon.
In such a way, pairing correlations, continuum, deformation, blocking effects,
and extended spatial density distributions in exotic, odd-$A$ or odd-odd nuclei can be
taken into account microscopically and self-consistently.

In order to treat pairing correlations, the quasi-particle concept is adopted
and the ground state of an even-even nucleus $|\Phi\rangle$ is represented as
a vacuum with respect to quasi-particles~\cite{Ring1980},
\begin{eqnarray}
 \beta_k | \Phi_0 \rangle = 0 , \quad \textrm{for all } k = 1, \ldots N ,
\end{eqnarray}
where $N$ is the dimension of the quasi-particle space and the quasi-particle
operators $\beta^\dag_k, \beta^{}_k$ are obtained by the Bogoliubov
transformation from the particle operators $c^\dag_l, c^{}_l$,
\begin{eqnarray}
  \left( \beta \atop \beta^\dag \right)
   = W^\dag  \left( c \atop c^\dag \right) , \quad
  W = \left( {U \atop V} {V^* \atop U^*} \right) .
\end{eqnarray}

The matrix $W$ of the coefficients $U$ and $V$ is unitary which guarantees that
the quasi-particle operators satisfy the same anti-communication rules as
the particle operators~\cite{Ring1980}.
Starting from the bare vacuum $|0\rangle$, the ground state of a system with
an even number of particles $| \Phi_0 \rangle$ can be constructed as,
\begin{eqnarray}
 | \Phi_0 \rangle = \prod_k \beta_k | 0\rangle,
\end{eqnarray}
where $k = 1, \ldots N$.
For an odd system, in practice, the corresponding ground state can be
constructed as one quasi-particle state $| \Phi_1 \rangle$,
\begin{eqnarray}
 | \Phi_1 \rangle = \beta^\dag_1 | \Phi_0 \rangle= \beta^\dag_1 \prod_{k} \beta_k | 0\rangle ,
\end{eqnarray}
where $\beta_1^\dag$ corresponds to the quasi-particle state with the lowest
quasi-particle energy.
In other words, the one quasi-particle state $| \Phi_1 \rangle$ is the vacuum
with respect to the set of quasi-particle operators $(\beta'_1, \ldots, \beta'_N)$ with
\begin{eqnarray}
 \beta'_1 = \beta^\dag_1, \
 \beta'_2 = \beta^{}_2, \ \ldots , \
 \beta'_N = \beta^{}_N ,
\end{eqnarray}
and the exchange of the operators $\beta^\dag_1 \leftrightarrow \beta^{}_1$ forms
a new set of quasi-particle operators $(\beta'_1, \ldots, \beta'_N,\ \beta'^\dag_1,
\ldots, \beta'^\dag_N)$, which corresponds to the exchange of the columns
$( U^{}_{l1}, V^{}_{l1} ) \longleftrightarrow ( V^*_{l1}, U^*_{l1} ) $ in the matrix $W$.
That is, the blocking effect in the odd system can be realized by exchanging the creator
$\beta^\dag_1$ with the corresponding annihilator $\beta^{}_1$ in the quasi-particle space.
Accordingly, the blocking effect in a multi-quasi-particle configurations can be treated.

The covariant density functional theory has been applied to describe successfully
nuclear structure over the entire periodic table~\cite{Serot1986_ANP16-1,
Reinhard1989_RPP52-439, Ring1996_PPNP37-193,Vretenar2005_PR409-101, Meng2006_PPNP57-470}.
The relativistic Hartree Bogoliubov equations~\cite{Ring1996_PPNP37-193,
Vretenar2005_PR409-101, Meng2006_PPNP57-470} for the nucleons read,
\begin{eqnarray}
 \int d^3 \bm{r}'
 \left(
  \begin{array}{cc}
   h_D
   - \lambda &
   \Delta
   \\
  -\Delta^*
   & -h_D
   + \lambda \\
  \end{array}
 \right)
 \left(
  { U_{k}
  \atop V_{k}
   }
 \right)
 & = &
 E_{k}
  \left(
   { U_{k}
   \atop V_{k}
    }
  \right)
 ,
 \label{eq:RHB0}
\end{eqnarray}
where $E_{k}$ is the quasiparticle energy, $\lambda$ is the chemical potential
which guarantees the proper average particle number and $h_D$ is the Dirac Hamiltonian,
\begin{equation}
 h_D(\bm{r}, \bm{r}') =
  \bm{\alpha} \cdot \bm{p} + V(\bm{r}) + \beta (M + S(\bm{r})).
\label{eq:Dirac0}
\end{equation}
The scalar and vector potentials
\begin{eqnarray}
 S(\bm{r}) & = & g_\sigma \sigma(\bm{r}), \label{eq:vaspot} \\
 V(\bm{r}) & = & g_\omega \omega^0(\bm{r}) +g_\rho \tau_3 \rho^0(\bm{r})
                +e \displaystyle\frac{1-\tau_3}{2} A^0(\bm{r}) ,
\label{eq:vavpot}
\end{eqnarray}
depend on the scalar field $\sigma$ as well as the time-like components
$\omega^0$, $\rho^0$, and $A^0$ of the iso-scalar vector field $\omega$,
the 3-component of iso-vector vector field $\rho$ and the photon field $A$.

For a fully paired and axially symmetric deformed system with the time reversal symmetry,
the projection of the total angular momentum on the symmetry axis $\Omega$
is conserved and each single particle state has a degeneracy of two.
The RHB equation~(\ref{eq:Dirac0}) can be reduced to half dimension $M=N/2$ and
can be decomposed into degenerate blocks with quantum numbers $+\Omega$ or $-\Omega$.
The corresponding density and abnormal density matrix have dimension $M$ and read
\begin{eqnarray}
  \rho_{ _{M \times M} }   &=& V^* V^T , \\
  \kappa_{ _{M \times M} } &=& V^* U^{T} ,
\end{eqnarray}
where $V$ and $U$ are the coefficients in the corresponding subspace.

For an odd system with the $k_b$-th level blocked in the $+\Omega$ subspace,
the time reversal symmetry is violated and there appear currents in the system.
These currents show an axial symmetry, i.e., $\Omega$ remains a good quantum number,
but the quasi-particle energies are no longer degenerate for the two subspaces,
because the subspace with $+\Omega$ contains the odd particle and
the corresponding subspace with $-\Omega$ contains an empty level.
Therefore, in principle, we have to diagonalize the RHB equation twice,
one for the subspace with $+\Omega$ and the other for the subspace with $-\Omega$.
Using the equal filling approximation which is usually made~\cite{Schunck2010_PRC81-024316,
Perez-Martin2008_PRC78-014304},
we neglect currents and average in a statistical manner over the two configurations
of a particle in the $+\Omega$
space and a particle in the $-\Omega$ space.
The corresponding currents cancel each other and in this way we obtain
in each step of the iteration fields with the time reversal symmetry.
In practice we average the density matrix $\rho$ and symmetrize the abnormal
density $\kappa$ in two subspaces and replace the two densities by
\begin{eqnarray}
  \rho' & = & \rho_{ _{M \times M} }
         + \frac{1}{2}
           \left( U_{k_b} U^{*T}_{k_b} -  V^*_{k_b} V^T_{k_b} \right) ,
    \label{eq:rhodd} \\
  \kappa' & = & \kappa_{ _{M \times M} }
         -  \frac{1}{2} \left( U_{k_b} V^{*T}_{k_b} + V^*_{k_b} U^T_{k_b} \right) ,
    \label{eq:kapodd}
\end{eqnarray}
where $V_{k_b}$ and $U_{k_b}$ are column vectors in the matrices $V$ and $U$
corresponding to the blocked level.
Note that Eqs.~(\ref{eq:rhodd}) and (\ref{eq:kapodd}) are the same as those
given in Refs.~\cite{Schunck2010_PRC81-024316, Perez-Martin2008_PRC78-014304}
if one considers the time reversal
symmetry and that the dimension of the densities here
is $M = N/2$.

\begin{table}
\begin{center}
\caption{\label{tab:O19}
Ground state properties of $^{19}$O from deformed RHB and RCHB calculations.
$R_{\rm n}$, $R_{\rm p}$ and $R_{\rm t}$ refer to neutron, proton and
total root mean square radii in unit of ``fm''.
$E_{\rm part}$, $E_\sigma$, $E_\sigma^{\rm non}$, $E_\omega$, $E_\omega^{\rm non}$, $E_\rho$,
$E_{\rm coul}$, $E_{\rm pair}$ and $E_{\rm tot}$ refer to energies of particle,
$\sigma$ field, non-linear term of $\sigma$ field, $\omega$ field,
non-linear term of $\omega$ field, $\rho$ field, Coulomb field,
paring energy and total binding energy in unit of ``MeV''.
}
\begin{tabular}{l | rr }
\hline\hline
  & deformed RHB &  RCHB  \\
\hline
  $R_{\rm n}         $ &  $    2.83$  &  $    2.83$    \\
  $R_{\rm p}         $ &  $    2.57$  &  $    2.57$    \\
  $R_{\rm t}         $ &  $    2.72$  &  $    2.72$    \\
  $E_{\rm part}      $ &  $ -445.60$  &  $ -445.73$    \\
  $E_\sigma          $ &  $ 2259.26$  &  $ 2260.18$    \\
  $E_\sigma^{\rm non}$ &  $  -77.14$  &  $  -77.18$    \\
  $E_\omega          $ &  $-1880.10$  &  $-1880.83$    \\
  $E_\omega^{\rm non}$ &  $   32.76$  &  $   32.77$    \\
  $E_\rho            $ &  $   -3.15$  &  $   -3.16$    \\
  $E_{\rm coul}      $ &  $  -17.10$  &  $  -17.10$    \\
  $E_{\rm pair}      $ &  $   -3.82$  &  $   -3.82$    \\
  $E_{\rm tot}       $ &  $ -144.80$  &  $ -144.78$    \\
\hline
\end{tabular}
\end{center}
\end{table}

\begin{figure}
\begin{center}
\includegraphics[width=0.21\textwidth]{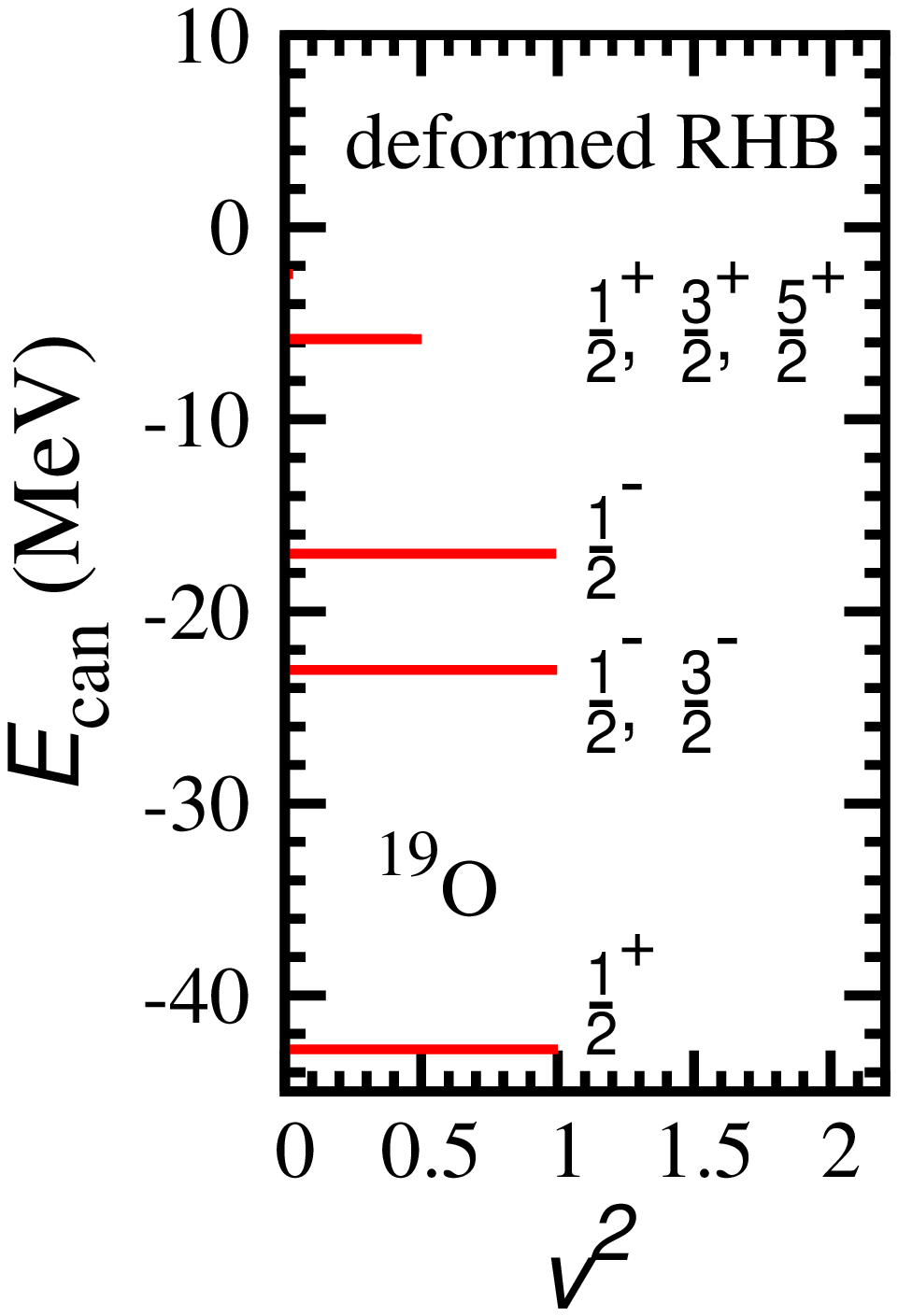}
~~~
\includegraphics[width=0.21\textwidth]{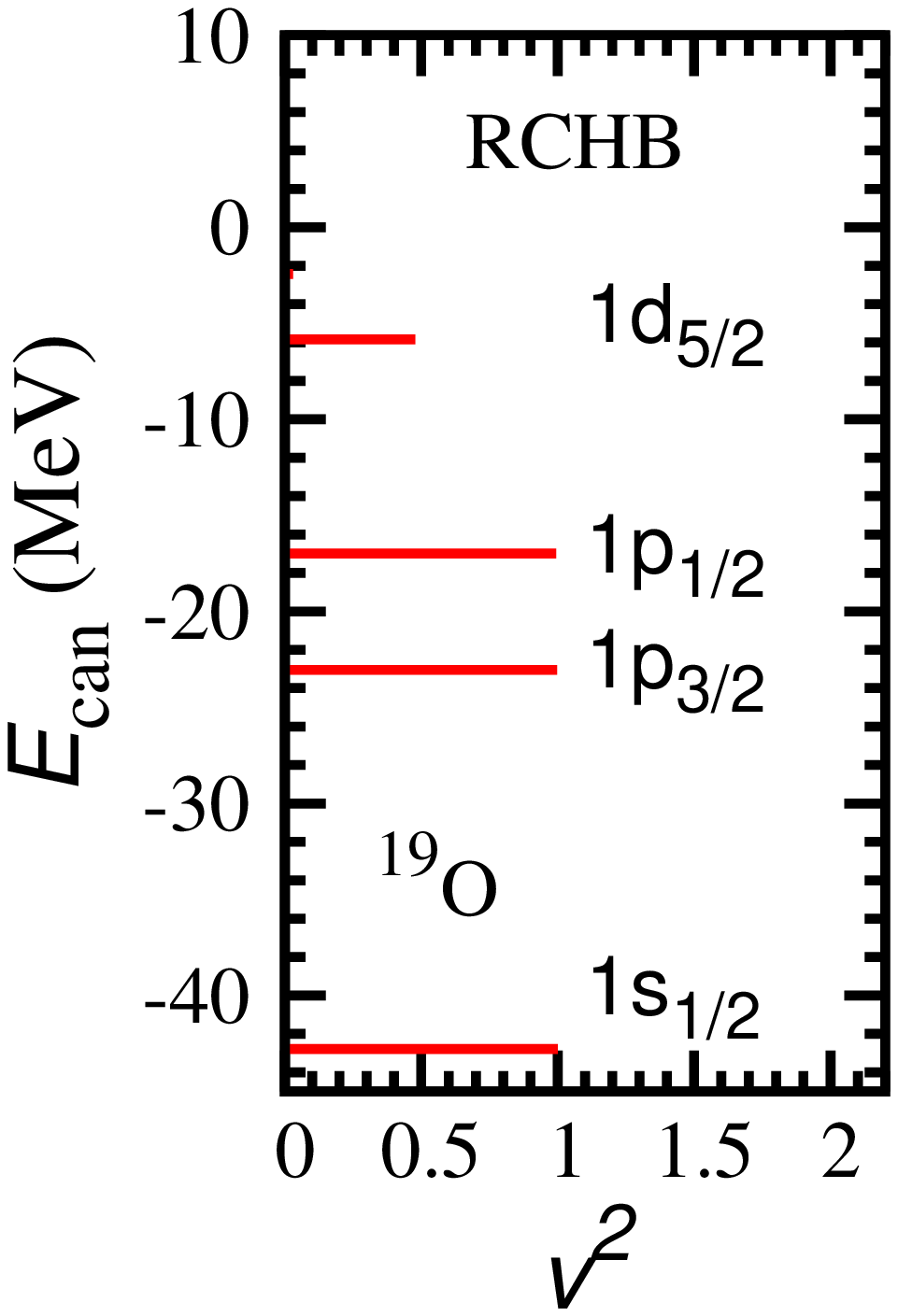}
\end{center}
\caption{(Color online)
\label{fig:O19_lev}
Neutron single particle levels in the canonical basis and their occupation
probability $v^2$ for $^{19}$O from the deformed RHB and RCHB calculations.
}
\end{figure}

In the following applications we use the density functional PK1~\cite{Long2004_PRC69-034319}
and a zero range density dependent pairing force~\cite{Meng1998_NPA635-3, Li2012_PRC85-024312}.
The deformed RHB equations are solved in a spherical Dirac Woods-Saxon
basis~\cite{Zhou2003_PRC68-034323} and the deformed potentials and densities are
expanded in terms of the Legendre polynomials $P_\lambda(\cos \theta)$
(for details see Refs.~\cite{Zhou2008_ISPUN2007, Zhou2010_PRC82-011301R, Li2012_PRC85-024312}).
In order to check the accuracy of the deformed code with the blocking effect
we investigate the spherical nucleus $^{19}$O and compare the results of
the deformed RHB code allowing only the spherical components of the fields,
i.e., $\lambda = 0$, with those obtained in the spherical relativistic continuum
Hartree Bogoliubov (RCHB) theory~\cite{Meng1998_NPA635-3}.
The ground state properties from the deformed RHB and the RCHB calculations
are given in Table~\ref{tab:O19}.
It is clearly seen that both calculations agree well with each other.
For example, the difference between the total binding energies is about 0.02 MeV,
which corresponds to an accuracy of about 0.01\%.

In order to check the details of the deformed RHB calculation,
we show in Fig.~\ref{fig:O19_lev} the neutron single particle levels
in the canonical basis for $^{19}$O in comparison with the spherical RCHB results.
The length of each level is proportional to the occupation probability.
In the spherical RCHB calculation, the blocked orbital is 1d$_{5/2}$.
In the deformed RHB code, a spherical solution is enforced by allowing only
$\lambda=0$ component in the Legendre expansion mentioned earlier
and therefore the three sublevels of the 1d$_{5/2}$ orbital with
$\Omega^{\pi} ={1}/{2}^+$, ${3}/{2}^+$, and ${5}/{2}^+$ are degenerate.
The present results in Fig.~\ref{fig:O19_lev} are obtained with the level
$\Omega^{\pi} = {1}/{2}^+$ blocked in the deformed RHB code,
but blocking any of these three levels gives the same results.
From Fig.~\ref{fig:O19_lev} one finds a good agreement between the results
from the deformed RHB and RCHB calculations.

\begin{figure}
\begin{center}
\includegraphics[width=0.4\textwidth]{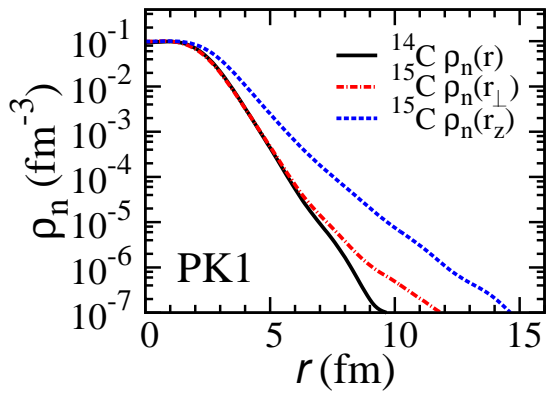}
\end{center}
\caption{(Color online) \label{fig:C15}
Neutron density profiles of $^{14}$C and $^{15}$C in deformed RHB calculations
with the parameter set PK1.
}
\end{figure}

\begin{figure}
\begin{center}
\includegraphics[width=0.21\textwidth]{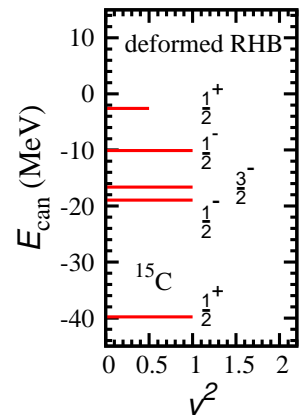}
\end{center}
\caption{(Color online)
\label{fig:C15_lev}
Neutron single particle levels in the canonical basis and their occupation
probability $v^2$ for $^{15}$C in the deformed RHB calculation.
}
\end{figure}

After checking the code in the spherical case we now turn to an application of
the deformed RHB theory for an exotic and deformed nucleus with an odd number of particles.
The deformed halo candidate nucleus $^{15}$C~\cite{Fang2004_PRC69-034613} is studied
with the parameter set PK1~\cite{Long2004_PRC69-034319}.
In contrary to the spherical even-even nucleus $^{14}$C, the deformed
RHB theory predicts a deformation $\beta = 0.25$ for the odd-$A$ nucleus $^{15}$C.
The neutron root mean square radii of $^{14}$C and $^{15}$C are 2.56 fm and 2.79 fm respectively.

The deformed RHB calculations lead for $^{15}$C to a deformed density
distributions with an axial symmetry.
In Fig.~\ref{fig:C15} the neutron densities along the symmetry axis
$\rho_n(r_z, r_\perp = 0)$ and perpendicular to the symmetry axis
$\rho_n(r_z=0, r_\perp)$ are plotted as dashed and dashed-dotted lines.
We also include the spherical density for $^{14}$C as a reference by a solid line.
It is interesting to see that, in the direction perpendicular to
the symmetry axis, the neutron densities are almost the same for $^{14}$C
and $^{15}$C at least when $r_\perp < 6$ fm.
Along the symmetry axis the neutron density of $^{15}$C extends much further
than that of $^{14}$C.
This is partly due to that $^{15}$C is prolate and the weakly bound
${1}/{2}^+$ level is occupied.

The single particle levels of $^{15}$C in the canonical basis are plotted in Fig.~\ref{fig:C15_lev}.
Since it is a deformed nucleus, the 1p$_{3/2}$ orbit is split into two levels with
$\Omega^{\pi} ={1}/{2}^-$ and $\Omega^{\pi} ={3}/{2}^-$ respectively.
There is one neutron occupying in $\Omega^{\pi} ={1}/{2}^+$ level near the threshold.
The occupation probability $v^2=0.5$, indicating that it is averaged over the two
configurations with $\Omega^{\pi}=\pm{1}/{2}^+$.
Because of the deformation, this level is a mixture of the spherical orbits
1d$_{5/2}$ (62\%) and 2s$_{1/2}$ (36\%).
The weakly-bound feature and the relatively large s-wave component of this level
results in that the neutron density of $^{15}$C extends further along the symmetry axis.

In summary, the blocking effect due to an odd nucleon is incorporated in
a deformed relativistic Hartree Bogoliubov (RHB) theory in continuum in order to
describe odd-$A$ or odd-odd exotic nuclei.
The formalism is briefly presented and the numerical checks are carried out
by comparing the results of the deformed RHB code for the spherical nucleus
$^{19}$O with those obtained from the spherical RCHB code.
As a first application, the deformed nucleus $^{15}$C is studied
by examining the neutron levels, deformation as well as density distributions
along and perpendicular to the symmetry axis.
Along the symmetry
axis the neutron density of $^{15}$C extends much further
because $^{15}$C is prolate and the weakly bound valence
level with $\Omega^\pi = {1}/{2}^+$ has a relatively large s-wave
component.

\begin{acknowledgments}
This work has been supported in part by
the Natural Science Foundation of China
(10875157, 10975100, 10979066, 11105005, 11175002, and 11175252),
by the Major State Basic Research Development Program of China (2007CB815000),
by the Knowledge Innovation Project of Chinese Academy of Sciences
(KJCX2-EW-N01 and KJCX2-YW-N32),
by the Oversea Distinguished Professor Project from Ministry of Education (MS2010BJDX001),
and by the DFG cluster of excellence \textquotedblleft Origin and Structure of the
Universe\textquotedblright\ (www.universe-cluster.de).
The computation was supported by the Supercomputing Center, CNIC of CAS.
\end{acknowledgments}


%

\end{document}